\documentclass[letterpaper,journal]{IEEEtran}
\usepackage{amsmath,amsfonts,amssymb,bm,dsfont}
\usepackage{booktabs}
\usepackage{multirow}
\usepackage{threeparttable}
\usepackage{array}
\usepackage{graphicx}
\usepackage[table]{xcolor}
\usepackage{indentfirst}
\usepackage{stfloats}
\usepackage{txfonts}
\usepackage{makecell}
\usepackage{cite}
\usepackage{xurl}
\usepackage{hyperref}

\newcommand{\g}[1]{\textcolor{gray}{#1}}

\begin{document}

\title{SAIL: Spatial Audio Intelligence with Large Language Models via Disentangled Acoustic-Spatial Encoding and Dual-Stream Q-Former}

\author{Zhengding Luo,~\IEEEmembership{Graduate Student Member,~IEEE,} Jinyang Wu, Haozhe Ma, Yanghao Zhou, Woon-Seng Gan,~\IEEEmembership{Senior Member,~IEEE,} and Wenwu Wang,~\IEEEmembership{Fellow,~IEEE}
\thanks{Z. Luo and W.-S. Gan are with the School of Electrical and Electronic Engineering, Nanyang Technological University, Singapore. E-mail: \{luoz0021, ewsgan\}@e.ntu.edu.sg.

J. Wu is with Singapore Management University. E-mail: \nolinkurl{jinyang.wu.2024@msc.smu.edu.sg}. 

H. Ma is with Tencent Hy Frontier Lab, Singapore. E-mail: \nolinkurl{haozhe.ma@u.nus.edu}.

Y. Zhou is with the Department of Computer Science and Technology, Beijing Institute of Technology, China. E-mail: \nolinkurl{zhouyh77@bit.edu.cn}.

W. Wang is with the University of Surrey, UK. E-mail: \nolinkurl{W.Wang@surrey.ac.uk}.

The code will be available at \url{https://github.com/Luo-Zhengding/SAIL}}}


\maketitle

\begin{abstract}
Spatial audio large language models (LLMs) enable embodied agents, wearable assistants, and immersive systems to recognize sound events, localize sources, and reason about their spatial relationships. However, existing spatial audio LLMs often rely on early fusion of acoustic and spatial features and source-agnostic token representations. These designs make it difficult to preserve the correspondence between individual sound events and their spatial attributes, particularly in multi-source scenes. To address this limitation, we propose SAIL, a Spatial Audio Intelligence framework with LLMs that preserves acoustic-spatial structure and source-level correspondence from audio encoding to LLM alignment. SAIL introduces a Disentangled Spatial Audio Transformer that represents Mel-spectrogram and interaural phase difference features as separate acoustic and spatial streams. Source-discriminative task queries further learn event, direction, and distance information for each source. A Dual-Stream Q-Former then aligns the two streams with the LLM using acoustic and spatial queries organized by source slots. Compared with the early-fusion baseline, SAIL achieves consistent improvements in dual-source sound event detection, direction and distance estimation, and spatial reasoning. These results demonstrate the importance of structured, source-discriminative audio representations for multi-source spatial understanding and reasoning.

\end{abstract}


\section{Introduction}
Audio is a fundamental modality for understanding and interacting with the physical world. Beyond recognizing what sound is present, humans can naturally infer where a sound comes from, how far away it is, and how multiple sound sources are arranged in a three-dimensional environment. Such spatial hearing ability is essential for embodied agents, wearable assistants, augmented/virtual reality (AR/VR) systems, and robots operating in dynamic real-world scenes \cite{SpatialAudioGeneration, wearvox, SLAM-LLM}. For example, an intelligent assistant should be able to identify whether a speaker is addressing the device or talking to someone else, determine which sound source is closer in a noisy environment, and reason about the relative locations of multiple simultaneous events. These capabilities require spatial audio understanding and reasoning that goes beyond conventional audio classification or captioning \cite{listen-Think, AudioThinker, Dspast,Hearyouare}.

Recent large audio-language models (LALMs) have made substantial progress in general audio understanding and reasoning by connecting audio encoders with large language models (LLMs) \cite{salmonn, AudioFlamingo, AudioFlamingo2, Qwen2-Audio, Kimi-Audio, Gama}. These models can answer questions about speech, music, and environmental sounds, showing the potential of LLMs as a unified interface for auditory perception and reasoning \cite{mmau, almtokenizer}. However, most existing audio-language models (ALMs) are built on monaural or channel-collapsed audio representations, which largely discard spatial cues such as interaural phase differences, direction of arrival, and distance-related acoustic patterns \cite{AudioWorldSim}. As a result, although existing LALMs can recognize the semantic content of audio clips, they remain limited in perceiving, understanding, and reasoning about the 3D spatial information of acoustic scenes \cite{starbench, Owl, Sci-phi}.

Spatial audio LLMs have recently emerged to address this limitation. A representative work is BAT~\cite{bat}, which introduces SpatialSoundQA and integrates a binaural spatial audio encoder, spatial audio spectrum transformer (Spatial-AST), with an LLM for spatial sound question answering. BAT demonstrates that binaural audio can support perception tasks such as sound event detection, direction estimation, and distance prediction, as well as spatial reasoning over two sound sources. More recent works further highlight the importance of spatial audio cues in realistic scenarios, including spatial audio generation \cite{SpatialAudioGeneration}, multi-channel audio-language modeling \cite{spur,phasecoder}, and spatial-temporal audio reasoning \cite{starbench}. Collectively, these studies establish spatial audio as an important component of context-aware LALMs. However, incorporating spatial cues does not necessarily provide the structured source-level representations required for fine-grained multi-source reasoning.

Despite recent progress, constructing sufficiently structured representations for source-level spatial reasoning remains an open challenge. For example, in BAT~\cite{bat}, Mel-spectrogram and Interaural Phase Difference (IPD) features are fused at the early stage and encoded by Spatial-AST into a shared token sequence. The audio tokens are then mapped to the LLM as a homogeneous audio representation, with no explicit distinction between acoustic semantics and spatial cues. This early-fusion design makes the resulting representation ambiguous: the LLM has to implicitly infer which parts of the fused tokens describe sound content and which encode localization information. A second, separate limitation lies in the absence of source-discriminative structure. Source-level spatial reasoning requires binding each sound event to its own direction and distance before spatial relationships across sources can be compared. However, when cues from multiple sources are compressed into a shared, source-agnostic token stream, the correspondence between each source and its spatial attributes is not explicitly preserved. BAT therefore faces two distinct challenges: \textbf{acoustic and spatial cues are entangled} before LLM alignment, leaving spatial information only implicitly accessible; and it does not introduce a source-discriminative structure, as \textbf{single- and multi-source inputs are processed with the same token and query design}, making fine-grained reasoning over multiple sound sources difficult.

\begin{figure}[!tp]
\centering
\includegraphics[width=\linewidth]{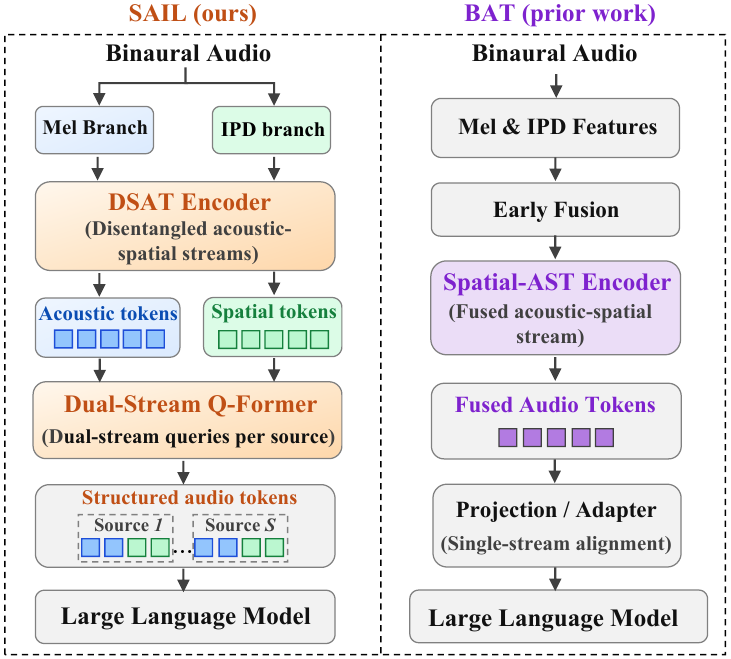}
\caption{Comparison between the proposed SAIL model and the prior BAT model. BAT relies on early-fused acoustic-spatial features and single-stream audio-language alignment, whereas SAIL disentangles acoustic and spatial cues and employs a Dual-Stream Q-Former to generate structured audio tokens.}
\label{Fig Compare SAIL with BAT}
\end{figure}

To address these limitations, we propose \textbf{SAIL}, a spatial audio intelligence framework with LLMs that introduces structured acoustic-spatial representations from audio encoding to LLM alignment. The comparison between the proposed SAIL model and the BAT model is illustrated in Fig.~\ref{Fig Compare SAIL with BAT}. At the encoder level, SAIL employs a \textbf{Disentangled Spatial Audio Transformer (DSAT)} to model Mel-spectrogram and IPD features as separate acoustic and spatial token streams, preserving their distinct roles while enabling token-level joint modeling through a shared Transformer. To further support source-discriminative perception, DSAT uses source-discriminative task queries during pre-training, with each source slot assigned three dedicated queries for event classification, direction estimation, and distance prediction. At the projector level, SAIL employs a \textbf{Dual-stream Q-Former} that uses acoustic and spatial queries to separately attend to the corresponding streams, with each query group further organized by source slots. This yields source-discriminative acoustic and spatial tokens, enabling the LLM to access compact representations that explicitly separate sound semantics from spatial cues while preserving their source-level correspondence. Consequently, SAIL can reason about what is heard, where each source is located, and how different sources are spatially related.

The proposed SAIL model is evaluated on single-source and dual-source spatial audio perception tasks, including sound event detection, direction-of-arrival estimation, and distance prediction, as well as dual-source spatial reasoning tasks involving direction-conditioned source identification, relative direction reasoning, and inter-source distance reasoning. Compared with the early-fusion baseline BAT, SAIL maintains comparable single-source perception while achieving better performance on dual-source perception and most spatial reasoning tasks. Additional experiments examine the compatibility of SAIL with different LLM backbones and compare it with general-purpose LALMs. These results highlight the importance of preserving acoustic-spatial structure and source-level correspondence throughout the encoder, projector, and LLM interface. Our contributions include:

\begin{itemize}
\item A Disentangled Spatial Audio Transformer (DSAT) that processes Mel-spectrogram and IPD features as separate acoustic and spatial token streams, avoiding early fusion of sound content and localization cues.
\item Source-discriminative task queries that provide structured source-level supervision during DSAT pre-training, enabling separate prediction of the sound event, direction, and distance associated with each source.
\item A Dual-Stream Q-Former Projector that uses acoustic and spatial queries organized by source slots to separately compress the two token streams and produce structured audio tokens for LLM alignment.
\item Compared with the early-fusion baseline, SAIL maintains comparable single-source perception while consistently improving two-source sound event detection, direction and distance estimation, and spatial reasoning.
\end{itemize}

\section{Related Work}
\subsection{Spatial Audio Perception}
Spatial audio perception is commonly studied through Sound Event Localization and Detection (SELD), where models aim to jointly identify sound events and estimate their spatial locations \cite{SELDnet, SALSA}. Related tasks such as direction-of-arrival (DoA) estimation and distance prediction further provide the basic ``what'' and ``where'' abilities required for spatial scene understanding. Existing SELD-style methods typically learn from handcrafted spatial cues or neural audio representations, and recent Transformer-based encoders further improve perception by jointly modeling spectral and spatial features \cite{PSELDNets,SelectTSL}. However, these methods are usually designed for closed-set and task-specific prediction, rather than open-ended language interaction or spatial question answering \cite{WorldNotMono, DynamicSoundSources}. In this work, we build upon this line of SELD-style spatial perception, but further connect the encoder to an LLM and introduce disentangled acoustic-spatial representations to support spatial reasoning over sound sources.

\begin{figure*}[!t]
\centering
\includegraphics[width=0.75\linewidth]{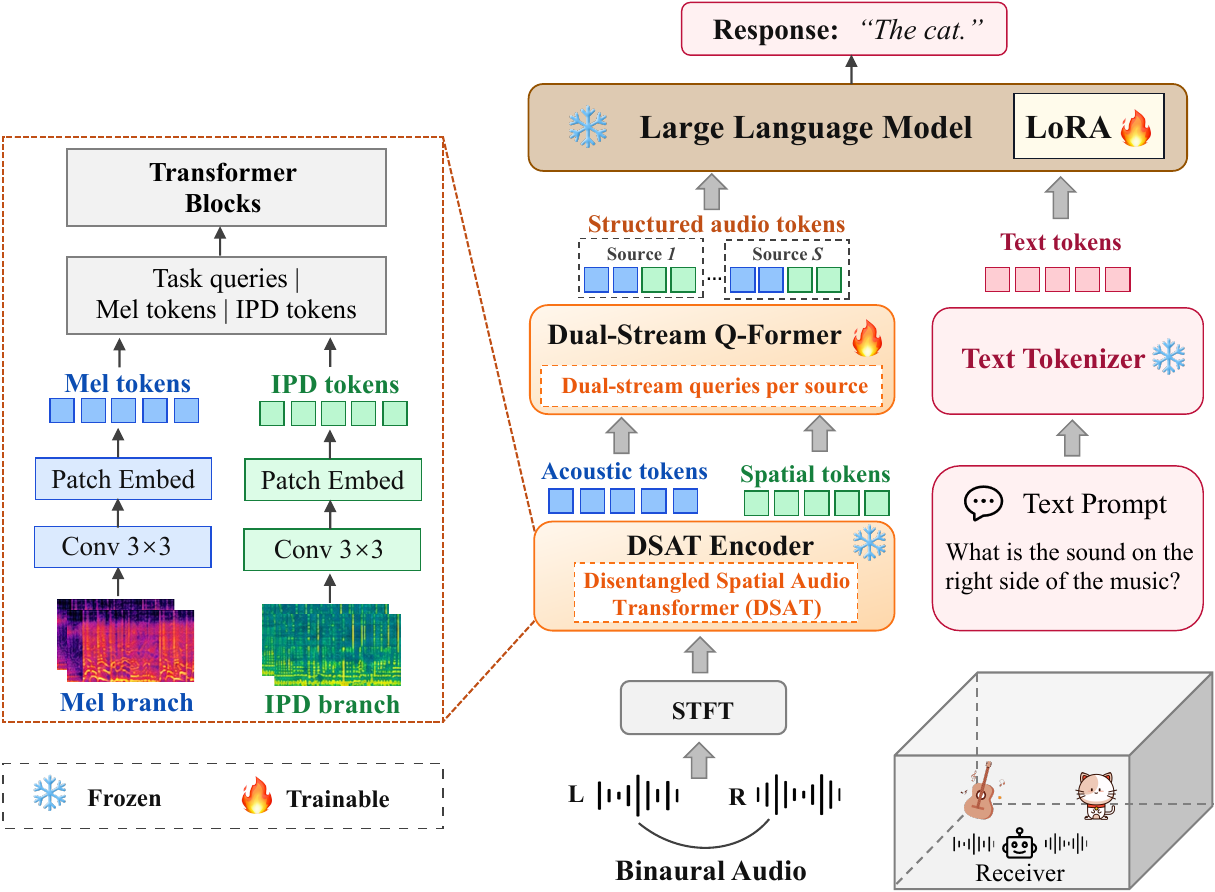}
\caption{Illustration of the proposed SAIL architecture. The DSAT encoder disentangles binaural audio into Mel-based acoustic tokens and IPD-based spatial tokens, while the Dual-Stream Q-Former uses dual-stream queries to extract spatial-audio tokens for each sound source. The resulting audio tokens are combined with text tokens and fed into the LoRA-adapted LLM for spatial audio-language reasoning.}
\label{Fig Main SAIL}
\end{figure*}

\subsection{Spatial Audio Large Language Models}
Large audio-language models have recently extended LLMs to audio understanding and reasoning. Many existing models, however, operate on monaural or channel-collapsed audio, limiting their ability to capture direction, distance, and inter-source spatial relationships~\cite{wearvox,starbench}. Spatial audio LLMs have therefore emerged to bridge the gap between audio-language reasoning and 3D acoustic perception. BAT~\cite{bat} introduces the SpatialSoundQA dataset and integrates the binaural Spatial-AST encoder with LLaMA-2, demonstrating sound event detection, direction and distance estimation, and two-source spatial reasoning. DSpAST~\cite{Dspast} extends Spatial-AST with feature attention and task-specific branches for sound event detection, distance prediction, and DoA estimation, focusing on task-level representation disentanglement. OWL~\cite{Owl} introduces a geometry-aware encoder trained with auxiliary depth and room-impulse-response supervision, together with spatially grounded chain-of-thought reasoning. Beyond binaural input, SPUR~\cite{spur} injects First-Order Ambisonics cues into existing LALMs through a lightweight spatial adapter, while PhaseCoder~\cite{phasecoder} learns spatial audio tokens from raw multichannel audio and microphone coordinates for geometry generalization.

Despite this progress, existing spatial audio LLMs still lack sufficiently structured representations for source-level spatial reasoning. For example, BAT relies on early fusion of acoustic and spatial cues in the encoder, which mixes sound content and localization information before LLM alignment. Moreover, it adopts a source-agnostic token and projection structure, where information from multiple sources is compressed into a shared token and query structure without source-discriminative slots. These two limitations make it difficult to explicitly associate each sound event with its corresponding direction and distance, especially in multi-source scenes. Among different spatial audio formats, binaural audio provides a practical two-channel input that is more accessible than First-Order Ambisonics (FOA) or specialized microphone arrays. Motivated by this, our work focuses on structured acoustic-spatial modeling of binaural audio, aiming to preserve both disentangled acoustic-spatial cues and source-level correspondence for spatial ALMs.

\subsection{Query-based Multimodal Alignment}
Query-based multimodal alignment has become a common strategy for connecting pretrained encoders for specific modalities with LLMs. Representative methods such as Q-Former in BLIP-2~\cite{blip2} introduce learnable queries to extract compact visual tokens from frozen vision encoders before projecting them into the language model space. Similar projector designs have been adopted in audio-language models to reduce long audio token sequences into a smaller set of LLM-compatible embeddings \cite{Next-gpt}. These methods are effective for general modality alignment, but standard query-based projectors typically compress all modality features into a single, source-agnostic token stream. For spatial audio, particularly in multi-source scenes, this compression can obscure both the distinction between acoustic semantics and spatial cues and the correspondence between each sound source and its spatial attributes. Source-level spatial reasoning instead requires the projection module to preserve structured information about what each sound source is, where it is located, and how it is spatially related to other sources. This motivates a dual-stream query-based projector with queries organized by source slots, which separately summarizes acoustic and spatial token streams for each source before the LLM alignment.

\section{Methodology}

\subsection{SAIL Overall Architecture}
As illustrated in Fig.~\ref{Fig Main SAIL}, SAIL connects spatial audio perception with LLM reasoning through two key designs: disentangled acoustic-spatial encoding and source-discriminative query structure. Given a binaural signal, time-frequency representations are first extracted and fed into the DSAT encoder, which processes Mel-based features and IPD-based features in two separate branches. Guided by source-discriminative task queries during pre-training, the encoder learns acoustic and spatial tokens that capture sound content and binaural localization cues, respectively. To align these representations with the LLM, a Dual-stream Q-Former separately summarizes the acoustic and spatial token streams using stream-specific queries. For multi-source scenes, these queries are organized by source slots, producing a compact set of source-discriminative acoustic and spatial representations. The resulting structured audio tokens are projected into the LLM embedding space, combined with the text embeddings of the user prompt, and fed into an LLM with Low-Rank Adaptation (LoRA). This enables SAIL to jointly reason about what is heard, where each source is located, and how multiple sources are spatially related.

\subsection{DSAT Encoder}

\textbf{Front-end Feature Extraction.}
The DSAT serves as the spatial audio encoder in SAIL. Given a binaural waveform with left and right channels $x_i(n)$, $i \in \{L, R\}$, we first apply the Short-Time Fourier Transform (STFT):
\begin{equation}\label{eq:stft}
X_i(t,f) = \sum_{n=0}^{N-1} x_i(tH+n)\, w(n)\, \exp\!\left(-j\frac{2\pi f n}{N}\right),
\end{equation}
where $i \in \{L, R\}, w(n)$ is the analysis window, $H$ the hop size, and $t, f$ index the time frame and frequency bin. Two complementary feature types are then extracted from the STFT representation. The log-Mel spectrogram of each channel captures acoustic semantics:
\begin{equation}
M_i(t,m) = \log\!\left( \sum_{f=1}^{F} |X_i(t,f)|\, \mathbf{W}_{\mathrm{mel}}(f,m) + \epsilon \right),
\end{equation}
where $\mathbf{W}_{\mathrm{mel}} \in \mathbb{R}^{F \times N_m}$ is the Mel filter-bank mapping $F$ linear-frequency bins to $N_m$ Mel bins, and $\epsilon$ ensures numerical stability. The IPD feature, which carries binaural spatial cues, is obtained from the phase spectra of the two channels:
\begin{equation}
\Phi(t,f) = \angle X_R(t,f) - \angle X_L(t,f).
\end{equation}
To remove phase-wraparound ambiguity, the IPD is decomposed into cosine and sine components and projected onto the Mel-frequency axis:
\begin{equation}
\begin{aligned}
P^{c}(t,m) &= \sum_{f=1}^{F} \cos\!\big(\Phi(t,f)\big)\, \mathbf{W}_{\mathrm{mel}}(f,m), \\
P^{s}(t,m) &= \sum_{f=1}^{F} \sin\!\big(\Phi(t,f)\big)\, \mathbf{W}_{\mathrm{mel}}(f,m).
\end{aligned}
\end{equation}
After Mel projection, the acoustic and spatial features share the same time-frequency resolution, which facilitates subsequent tokenization.

\textbf{Disentangled Spatial Audio Transformer.}
In contrast to Spatial-AST~\cite{bat}, which concatenates Mel and IPD features into a single multi-channel input, the DSAT encoder processes them in two parallel branches with no cross-branch interaction until the Transformer stage, preserving the semantic separation between \textit{what} and \textit{where} at the feature level. Early fusion in Spatial-AST entangles acoustic semantics and spatial cues at the lower representational level, making it harder for downstream encoder and projector modules to maintain their distinct roles. By deferring acoustic-spatial interaction to the self-attention layers, DSAT preserves a structured token layout for the downstream Dual-Stream Q-Former, which aligns the two streams with the LLM through separate query paths.

The Mel branch stacks the left and right log-Mel spectrograms into a two-channel acoustic tensor, while the IPD branch stacks the cosine- and sine-transformed IPD features into a two-channel spatial tensor:
\begin{equation}
\begin{aligned}
\mathbf{F}_{\mathrm{mel}} &= [M_L;\, M_R] \in \mathbb{R}^{2 \times T \times N_m}, \\
\mathbf{F}_{\mathrm{ipd}} &= [P^{c};\, P^{s}] \in \mathbb{R}^{2 \times T \times N_m},
\end{aligned}
\end{equation}
where $T$ is the number of time frames. Each branch begins with a 2D convolution followed by batch normalization and Gaussian Error Linear Unit (GELU) based activation, fusing local inter-channel information within each feature type:
\begin{equation}
\widetilde{\mathbf{F}}_{b} = \mathrm{GELU}\!\left( \mathrm{BN}\!\left( \mathrm{Conv}^{3\times 3}_{b}\!\left(\mathbf{F}_{b}\right) \right) \right), \quad b \in \{\mathrm{mel},\, \mathrm{ipd}\}.
\end{equation}
The convolution collapses the two input channels into a single feature map, which is then partitioned into patch tokens by a branch-specific patch-embedding layer:
\begin{equation}
\mathbf{Z}_{b} = \mathrm{PatchEmbed}_{b}\!\left( \widetilde{\mathbf{F}}_{b} \right) \in \mathbb{R}^{N_p \times d}, \quad b \in \{\mathrm{mel},\, \mathrm{ipd}\},
\end{equation}
where $N_p$ is the number of patches and $d$ the token dimension.

To guide the encoder toward source-discriminative spatial representations, we introduce learnable task queries during encoder pre-training. Given $S$ candidate sound sources, each source is assigned three task queries, i.e., target distance, DoA, and sound-event classes, respectively:
\begin{equation}\label{eq:queries}
\mathbf{Q}_{\mathrm{task}} = \big[\mathbf{q}^{(1)}_{\mathrm{dis}}, \mathbf{q}^{(1)}_{\mathrm{doa}}, \mathbf{q}^{(1)}_{\mathrm{cls}}, \ldots, \mathbf{q}^{(S)}_{\mathrm{dis}}, \mathbf{q}^{(S)}_{\mathrm{doa}}, \mathbf{q}^{(S)}_{\mathrm{cls}} \big] \in \mathbb{R}^{3S \times d}.
\end{equation}
The Transformer input is the concatenation of task queries, Mel tokens, and IPD tokens:
\begin{equation}
\mathbf{H}^{0} = \big[\, \mathbf{Q}_{\mathrm{task}};\, \mathbf{Z}_{\mathrm{mel}};\, \mathbf{Z}_{\mathrm{ipd}} \,\big] \in \mathbb{R}^{(3S+2N_p) \times d}.
\end{equation}
A stack of $N_T$ Transformer blocks is then applied:
\begin{equation}
\mathbf{H}^{\ell} = \mathrm{TransformerBlock}^{\ell}\!\left( \mathbf{H}^{\ell-1} \right), \quad \ell = 1, \ldots, N_T,
\end{equation}
yielding an output that preserves the input token layout:
\begin{equation}
\mathbf{H}^{N_T} = \big[\, \mathbf{T}_{\mathrm{task}};\, \mathbf{T}_{\mathrm{acoustic}};\, \mathbf{T}_{\mathrm{spatial}} \,\big],
\end{equation}
with $\mathbf{T}_{\mathrm{task}} \in \mathbb{R}^{3S \times d}$ and $\mathbf{T}_{\mathrm{acoustic}}, \mathbf{T}_{\mathrm{spatial}} \in \mathbb{R}^{N_p \times d}$. The task tokens are grouped by source slots,
\begin{equation}
\mathbf{T}_{\mathrm{task}} = \big[\, \mathbf{T}^{(1)}_{\mathrm{dis}},\, \mathbf{T}^{(1)}_{\mathrm{doa}},\, \mathbf{T}^{(1)}_{\mathrm{cls}},\, \ldots,\, \mathbf{T}^{(S)}_{\mathrm{dis}},\, \mathbf{T}^{(S)}_{\mathrm{doa}},\, \mathbf{T}^{(S)}_{\mathrm{cls}} \,\big],
\end{equation}
where $\mathbf{T}^{(s)}_{\mathrm{dis}}, \mathbf{T}^{(s)}_{\mathrm{doa}}, \mathbf{T}^{(s)}_{\mathrm{cls}} \in \mathbb{R}^{d}$ are the contextualized distance, DoA, and event tokens of the $s$-th source slot ($s = 1, \ldots, S$).

\begin{figure*}[!t]
\centering
\includegraphics[width=0.9\linewidth]{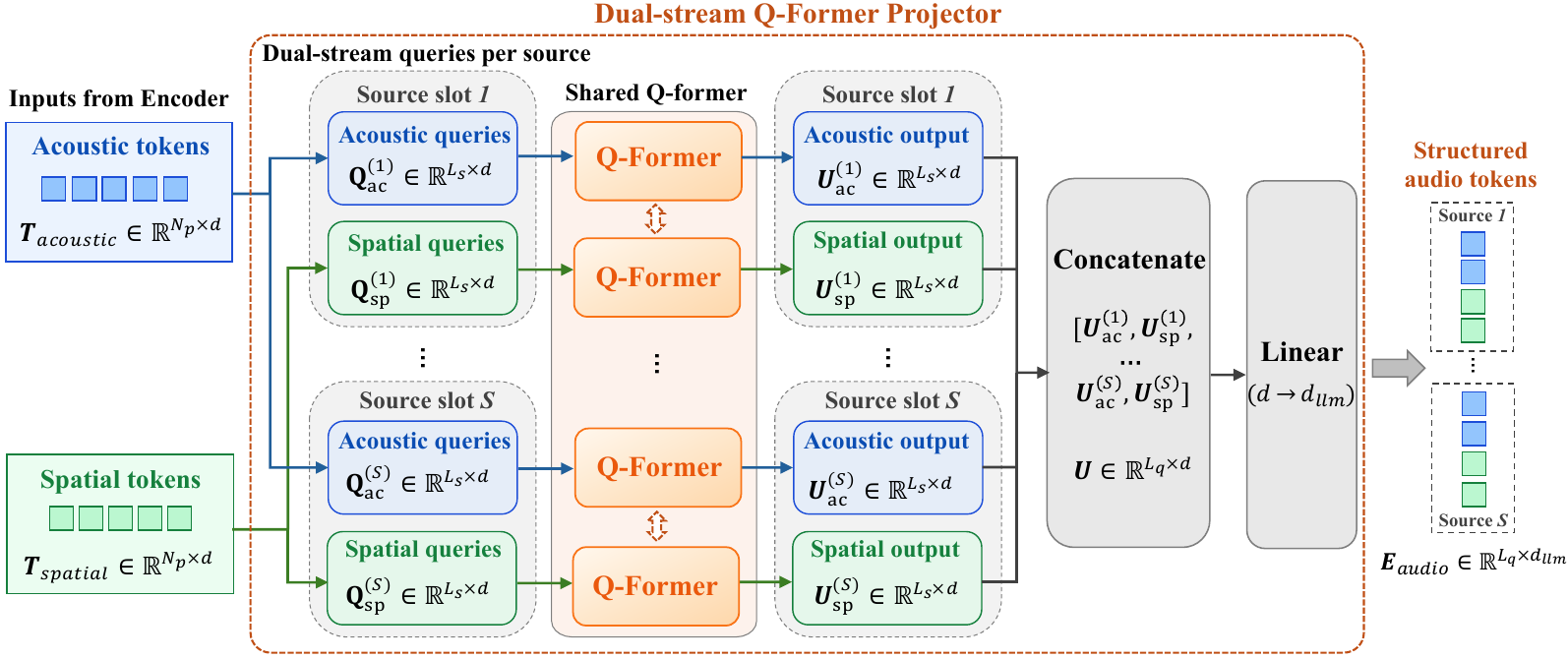}
\caption{Illustration of the Dual-Stream Q-Former Projector with $S$ source slots. Given the disentangled token streams from the DSAT encoder, acoustic and spatial queries cross-attend exclusively to their respective streams through a shared Q-Former, preserving the acoustic-spatial separation. The $2S$ resulting output groups are concatenated in source-major, stream-minor order and projected into LLM-compatible audio tokens.}
\label{Fig Dual-stream Q-Former}
\end{figure*}

\textbf{Multi-Task Pre-training Objective.}
During pre-training, the output task tokens are routed to task-specific prediction heads. For the $s$-th source slot,
\begin{equation}
\begin{aligned}
\hat{\mathbf{y}}^{(s)}_{\mathrm{cls}} &= g_{\mathrm{cls}}\!\left(\mathbf{T}^{(s)}_{\mathrm{cls}}\right),
&\hat{\mathbf{y}}^{(s)}_{\mathrm{dis}} &= g_{\mathrm{dis}}\!\left(\mathbf{T}^{(s)}_{\mathrm{dis}}\right),\\
\hat{\mathbf{y}}^{(s)}_{\mathrm{az}} &= g_{\mathrm{az}}\!\left(\mathbf{T}^{(s)}_{\mathrm{doa}}\right),
&\hat{\mathbf{y}}^{(s)}_{\mathrm{el}} &= g_{\mathrm{el}}\!\left(\mathbf{T}^{(s)}_{\mathrm{doa}}\right),
\end{aligned}
\end{equation}
where $g_{\mathrm{cls}}$, $g_{\mathrm{dis}}$, $g_{\mathrm{az}}$, and $g_{\mathrm{el}}$ are linear classifiers preceded by layer normalization, and the azimuth and elevation predictions are produced from a shared DoA token. Event classification is trained with a multi-label binary cross-entropy loss, whereas distance, azimuth, and elevation are discretized into bins and trained with cross-entropy, so that $\mathcal{L}_{\mathrm{doa}}$ is the average of the azimuth and elevation terms. The encoder is optimized with a multi-task objective:
\begin{equation}\label{loss encoder}
\mathcal{L}_{\mathrm{enc}} = \lambda_{\mathrm{cls}}\,\mathcal{L}_{\mathrm{cls}} + \lambda_{\mathrm{dis}}\,\mathcal{L}_{\mathrm{dis}} + \lambda_{\mathrm{doa}}\,\mathcal{L}_{\mathrm{doa}},
\end{equation}
where $\lambda_{\mathrm{cls}}$, $\lambda_{\mathrm{dis}}$, and $\lambda_{\mathrm{doa}}$ balance the three tasks. For single-source scenarios, only the first source slot is supervised, while the remaining slots do not receive a gradient. For multi-source cases, source slots are matched to ground-truth sources by permutation-invariant training~\cite{PIT, DETR-object}, where the matching cost is computed from the spatial losses (distance and DoA) only, as spatial attributes are physically unique to each source and provide a more stable matching signal than class labels, which may overlap across sources. The classification, distance, and DoA losses are then jointly back-propagated under the selected permutation.

\textbf{Source-discriminative Task Queries.}
The task queries $\mathbf{Q}_{\mathrm{task}}$ in Eq.~\ref{eq:queries} are introduced to provide source-level supervision during encoder pre-training. Since the patch-level Mel and IPD tokens represent the whole binaural scene, they do not explicitly indicate which sound event or spatial attribute belongs to which source. Inspired by the set-prediction formulation of DEtection TRansformer (DETR)~\cite{DETR-object}, the task queries serve as learnable source slots that aggregate source-specific information from the shared Mel and IPD token streams through self-attention, so that source separation is performed at the query level rather than by splitting the token streams themselves. Within each slot, three queries are assigned to sound event classification, distance prediction, and DoA estimation, which allows different queries to specialize toward different aspects of the Mel and IPD tokens. For multi-source scenes, the slots are trained without assuming a fixed source order, following the permutation-invariant matching introduced above. The task queries $\mathbf{Q}_{\mathrm{task}}$ remain part of the input sequence after pre-training, where they continue to shape the two branches through self-attention even though their output tokens are no longer read out. After encoder pre-training, the prediction heads and the output task tokens $\mathbf{T}_{\mathrm{task}}$ are discarded, and only $\mathbf{T}_{\mathrm{acoustic}}$ and $\mathbf{T}_{\mathrm{spatial}}$ are retained as \textbf{disentangled acoustic and spatial token streams}, which are then passed to the Dual-Stream Q-Former for alignment with the LLM as shown in Fig.~\ref{Fig Main SAIL}.

\subsection{Dual-Stream Q-Former Projector}
The DSAT encoder produces two disentangled token streams, namely acoustic tokens $\mathbf{T}_{\mathrm{acoustic}} \in \mathbb{R}^{N_p \times d}$ and spatial tokens $\mathbf{T}_{\mathrm{spatial}} \in \mathbb{R}^{N_p \times d}$, where $N_p$ is the number of patch tokens per stream and $d$ is the token dimension (the batch dimension is omitted for clarity). The Dual-stream Q-Former Projector compresses these two streams into a fixed-length LLM-compatible token sequence while maintaining their acoustic-spatial separation. Specifically, acoustic queries cross-attend only to acoustic tokens, whereas spatial queries cross-attend only to spatial tokens through a shared Q-Former. For multi-source scenes, the query groups are further organized by source slots, producing a structured token layout that preserves both stream-level and source-slot information before projection into the LLM embedding space. The architecture of the Dual-stream Q-Former Projector with $S$ source slots is illustrated in Fig.~\ref{Fig Dual-stream Q-Former}.

For each source slot $s \in \{1, \ldots, S\}$, we allocate two groups of learnable queries with $L_s$ queries per group:
\begin{equation}
\mathbf{Q}_{\mathrm{ac}}^{(s)},\ \mathbf{Q}_{\mathrm{sp}}^{(s)} \in \mathbb{R}^{L_s \times d}, 
\quad s = 1, \ldots, S.
\end{equation}
Here, $\mathbf{Q}_{\mathrm{ac}}^{(s)}$ and $\mathbf{Q}_{\mathrm{sp}}^{(s)}$ denote the acoustic and spatial query groups for source slot $s$, respectively. The total number of audio tokens fed into the LLM is $L_q = 2 S L_s$. A shared Q-Former $\mathcal{F}_{\mathrm{qf}}$, whose hidden dimension is also $d$, is reused across all source slots and streams. For $S$ source slots, it is invoked $2S$ times, corresponding to the acoustic and spatial streams of each source slot. Each invocation processes one query group in isolation and cross-attends only to its corresponding stream:
\begin{equation}
\begin{aligned}
\mathbf{U}_{\mathrm{ac}}^{(s)} &= \mathcal{F}_{\mathrm{qf}}\!\left( \mathbf{Q}_{\mathrm{ac}}^{(s)};\, \mathbf{T}_{\mathrm{acoustic}} \right), \\
\mathbf{U}_{\mathrm{sp}}^{(s)} &= \mathcal{F}_{\mathrm{qf}}\!\left( \mathbf{Q}_{\mathrm{sp}}^{(s)};\, \mathbf{T}_{\mathrm{spatial}} \right),
\end{aligned}
\end{equation}
yielding $\mathbf{U}_{\mathrm{ac}}^{(s)}, \mathbf{U}_{\mathrm{sp}}^{(s)} \in \mathbb{R}^{L_s \times d}$. The Q-Former therefore acts as a stream-specific token summarizer, compressing each acoustic or spatial token stream into a compact set of query outputs while preserving the structured source-stream layout.

The outputs from all source slots are concatenated in a source-major, stream-minor order:
\begin{equation}
\mathbf{U} = \big[\, \mathbf{U}_{\mathrm{ac}}^{(1)};\, \mathbf{U}_{\mathrm{sp}}^{(1)};\, 
\ldots;\, \mathbf{U}_{\mathrm{ac}}^{(S)};\, \mathbf{U}_{\mathrm{sp}}^{(S)} \,\big] 
\in \mathbb{R}^{L_q \times d}.
\end{equation}
A linear projection followed by layer normalization then maps $\mathbf{U}$ into the LLM embedding space:
\begin{equation}
\mathbf{E}_{\mathrm{audio}} = \mathrm{LayerNorm}\!\left( \mathrm{Linear}(\mathbf{U}) \right) 
\in \mathbb{R}^{L_q \times d_{\mathrm{llm}}},
\end{equation}
where $d_{\mathrm{llm}}$ is the hidden dimension of the LLM. The resulting $\mathbf{E}_{\mathrm{audio}}$ is a sequence of \textbf{structured audio tokens in the LLM embedding space}. These tokens are then combined with text embeddings and fed into the LLM for audio-language reasoning. The projector is formulated for a general number of source slots $S$. Each source-stream group is assigned a fixed number of $L_s$ queries, so that the per-source resolution remains unchanged as $S$ grows and the audio token budget $L_q = 2 S L_s$ scales linearly with the number of slots. Since the slots are ordered, a single-source scene is served by the same model and checkpoint by retaining only the $2 L_s$ tokens of the first slot, so that the single-source token sequence is a truncated prefix of the dual-source one rather than a separate architecture. This design improves the scalability of the SAIL model and facilitates its adaptation to acoustic scenes with varying numbers of sound sources.

\subsection{LLM-based Spatial Audio Reasoning} \label{LLM-based Spatial Audio Reasoning}
\textbf{Prompt Construction and Audio Embedding Insertion.} SAIL combines the disentangled audio embeddings $\mathbf{E}_{\mathrm{audio}}$ produced by the Dual-Stream Q-Former with a textual instruction through placeholder-based embedding replacement. A natural-language question is wrapped into an instruction-style prompt, and a contiguous span of $L_q$ placeholder tokens is prepended to it to reserve room for the audio input. The resulting token sequence is embedded into $\mathbf{E}_{\mathrm{text}} \in \mathbb{R}^{L_t \times d_{\mathrm{llm}}}$, where $L_t$ is its total length including the $L_q$ placeholder positions. At forward time, $\mathbf{E}_{\mathrm{audio}} \in \mathbb{R}^{L_q \times d_{\mathrm{llm}}}$ is written into the reserved placeholder positions. Let $\mathbf{E}_{\mathrm{audio}}^{\mathrm{pad}} \in \mathbb{R}^{L_t \times d_{\mathrm{llm}}}$ denote $\mathbf{E}_{\mathrm{audio}}$ zero-padded to length $L_t$, with its $L_q$ rows placed at the reserved placeholder positions and all other rows set to zero. A binary modality mask $\mathbf{M} \in \{0,1\}^{L_t \times 1}$ is used to indicate these placeholder positions. The final multimodal embedding sequence is computed as
\begin{equation}
\mathbf{E} = \mathbf{M} \odot \mathbf{E}_{\mathrm{audio}}^{\mathrm{pad}} 
           + (\mathbf{1}-\mathbf{M}) \odot \mathbf{E}_{\mathrm{text}},
\end{equation}
where $\odot$ denotes element-wise multiplication broadcast along the feature dimension. The placeholder embeddings are thereby replaced by audio embeddings while all remaining text embeddings are kept unchanged. The resulting multimodal sequence $\mathbf{E}$ is then fed into the LLM for autoregressive response generation.

\textbf{Parameter-Efficient Fine-tuning.}
Both the pre-trained DSAT encoder and the LLM backbone remain frozen throughout fine-tuning. Only two lightweight components are updated: (i) the Dual-Stream Q-Former, including the learnable queries, the shared $\mathcal{F}_{\mathrm{qf}}$ Transformer with cross-attention to the encoder tokens, and the final linear projection to the LLM embedding space; and (ii) LoRA adapters~\cite{lora} inserted into the self-attention projections of the LLM. This design isolates the burden of acoustic-spatial reasoning to the projector while preserving the linguistic capacity of the pre-trained LLM.

\textbf{Training Objective.}
SAIL is trained through a two-stage pipeline. In the first stage, the DSAT encoder is pre-trained with the multi-task objective $\mathcal{L}_{\mathrm{enc}}$ (Eq.~\ref{loss encoder}) in three sub-stages. The first two sub-stages use single-source data: the spatial losses are initially disabled so that the backbone and the event classification head converge on \emph{what} is in the scene, and are then activated on the converged backbone so that the encoder additionally learns \emph{where} each source is located. The third sub-stage introduces two-source mixtures, drawn with equal probability against single-source samples, and supervises both source slots under the permutation-invariant matching described above, so that the second slot is specialized while the first retains its single-source accuracy. Together, these sub-stages yield disentangled acoustic and spatial representations that generalize to both single-source and dual-source scenes.

In the second stage, the pre-trained encoder is integrated into SAIL and frozen, and the Dual-Stream Q-Former and LoRA adapters are jointly fine-tuned on the SpatialSoundQA dataset \cite{bat} with a standard language-modeling objective computed only over the response tokens. Following the perception-to-reasoning curriculum of BAT~\cite{bat}, fine-tuning proceeds through three sub-stages: (i) single-source perception, which teaches the model to identify the class, distance, and direction of an individual source; (ii) dual-source perception, which introduces mixed-source scenes and implicit source separation; and (iii) spatial reasoning, which extends the model to questions about relative spatial relationships between multiple sources while retaining the single-target questions of the previous sub-stage. This curriculum progressively exposes the LLM to increasing acoustic-spatial complexity while reusing the same set of trainable parameters throughout.

\section{Experiments}

\subsection{Dataset}
We build on the datasets introduced by BAT~\cite{bat} for both encoder pre-training and LLM fine-tuning. Spatial audio is synthesized by convolving monaural sound sources from AudioSet~\cite{AudioSet} with binaural room impulse responses (RIRs) generated by SoundSpaces~2.0~\cite{Soundspaces} in Matterport3D environments~\cite{Matterport3d}. Following BAT, we use the filtered AudioSet split with 355 sound-identifiable event categories, 1.86M training clips, and 17{,}148 evaluation clips. The acoustic paths are simulated over 90 real-world indoor scenes, resulting in 21{,}131 source-receiver binaural RIR instances. For encoder pre-training, we construct both single-source and two-source spatial audio for sound event classification, distance estimation, and DoA prediction. Unlike Spatial-AST in BAT, which is trained only on single-source audio with single-source task queries, the DSAT encoder uses source-discriminative task queries to learn from multi-source supervision and predict event, distance, and DoA for each source separately.

Following BAT~\cite{bat}, we use SpatialSoundQA to train the Dual-Stream Q-Former and the LLM's LoRA adapters and evaluate on the same official split. SpatialSoundQA contains roughly $872$K question-answering (QA) pairs organized into a three-stage perception-to-reasoning curriculum, with cumulative totals of approximately $278$K, $514$K, and $872$K QA pairs, respectively. Types A and B evaluate sound-event detection and DoA and distance prediction (DP) for single-source audio, respectively, while Types C and D evaluate the corresponding tasks for dual-source audio. Type E evaluates dual-source spatial reasoning through binary Yes/No questions about direction and distance and open-ended questions concerning direction-conditioned source identification, relative source direction, and inter-source distance.

\subsection{Model Implementation Details}

\textbf{Audio Pre-processing}. Each anechoic source waveform is resampled to $32$~kHz, normalized to $-14$~dB~FS based on its root-mean-square amplitude, and convolved with its corresponding binaural room impulse response. Multi-source scenes are constructed by averaging the convolved binaural signals, following the data construction protocol of BAT~\cite{bat}. Since convolution extends the signal with a reverberation tail, the resulting waveform is trimmed or zero-padded to $10$ seconds. For the STFT in Eq.~\eqref{eq:stft}, we use a window size of $N=1024$ and a hop size of $H=320$, and the Mel filter bank uses $N_m=128$ Mel bins. The resulting Mel and IPD features are aligned to a fixed length of $T=1024$ frames, giving $\mathbf{F}_{\mathrm{mel}} \in \mathbb{R}^{2\times 1024\times 128}$ and $\mathbf{F}_{\mathrm{ipd}} \in \mathbb{R}^{2\times 1024\times 128}$. Unlike Spatial-AST~\cite{bat}, which concatenates Mel and IPD features before tokenization, DSAT processes them separately through the Mel and IPD branches. The resulting tokens are concatenated with the task queries and fed into the shared Transformer.

\textbf{DSAT Encoder.} The Mel and IPD branches each first compress their two-channel inputs with a $3\times3$ convolution and then tokenize them using independent $16\times16$ patch embeddings, producing $512$ tokens per stream with an embedding dimension of $768$. The full sequence, consisting of $3S$ source-discriminative task queries, Mel tokens, and IPD tokens, is processed by a shared $12$-layer Transformer. The shared Transformer and compatible Mel-branch parameters are initialized from the official AudioMAE~\cite{AudioMAE} checkpoint. The Mel tokens use AudioMAE's frozen sinusoidal positional embeddings, whereas the IPD tokens use learnable positional embeddings. During pre-training, two-dimensional random token masking with a ratio of $0.25$ along each of the time and frequency axes is applied to both streams.

To stabilize multi-task optimization, DSAT pre-training proceeds in three phases. In the first phase, the single-source encoder is trained for $20$ epochs using only the event classification loss ($\lambda_{\mathrm{cls}}{:}\lambda_{\mathrm{dis}}{:}\lambda_{\mathrm{doa}}=1000{:}0{:}0$), with a base learning rate of $1\times10^{-3}$. In the second phase, it is trained for another $10$ epochs with all three losses enabled ($400{:}4{:}2$) and a base learning rate of $2\times10^{-4}$, allowing spatial prediction to be learned from the converged acoustic representations. In the third phase, the two-source DSAT encoder is initialized from the checkpoint of the second phase and trained for $20$ epochs on single-source and dual-source samples drawn with equal probability, using loss weights of $400{:}8{:}2$ and a base learning rate of $2\times10^{-4}$. Following AudioMAE~\cite{AudioMAE}, the actual learning rate is obtained by scaling the base learning rate by the effective batch size divided by $256$. Complete training configurations are provided in Appendix~\ref{Training Configurations}.

\textbf{Dual-Stream Q-Former Projector and LLM.}
The Dual-Stream Q-Former follows the Q-Former architecture~\cite{blip2}, with $8$ Transformer layers and a hidden dimension of $768$. The projector maintains $128$ learnable queries organized into two source slots, each containing $32$ acoustic and $32$ spatial queries. Although all queries are computed, only the $64$ outputs of the first slot are passed to the LLM for single-source inputs, while all $128$ outputs are used for dual-source inputs. The acoustic and spatial branches have independent query embeddings, while sharing the same Q-Former parameters. Their outputs are concatenated and mapped to the LLM embedding space through a linear layer followed by LayerNorm.

We use Llama-2-7B~\cite{Llama2} as the language backbone, with $d_{\mathrm{llm}}=4096$, following BAT~\cite{bat} for a controlled comparison. During multimodal instruction tuning, the DSAT encoder and the base LLM remain frozen, while the Dual-Stream Q-Former and LoRA adapters~\cite{lora} are optimized using the language modeling loss over response tokens. LoRA is applied to the query and value projections with rank $r=8$, scaling parameter $\alpha=32$, and dropout $0.05$. Training follows the three-stage curriculum described in Sec.~\ref{LLM-based Spatial Audio Reasoning}. Each stage is trained for $3$ epochs and initialized from the checkpoint of the preceding stage. We use AdamW with a peak learning rate of $2\times10^{-4}$. At each stage, the learning rate increases linearly during the first approximately $4.4\%$ of the optimization steps and then follows a linear decay schedule. The per-GPU batch size is $32$ for Stage I and $24$ for Stages II and III.

\begin{table*}[!t]
\centering
\caption{Comparison of spatial audio encoders on single-source and dual-source SELD tasks. "Dual-source: one source matched" scores each model only on its best-covered source, while "Dual-source: both sources matched" scores predictions for both sources after permutation matching. Metrics include mean Average Precision (mAP $\uparrow$), Error Rate at $20^{\circ}$ (ER$_{20^{\circ}}$ $\downarrow$), Mean Angular Error (MAE $\downarrow$), and Distance Error Rate (DER $\downarrow$).}
\begin{tabular*}{0.9\textwidth}{@{\extracolsep{\fill}}llccccc}
\toprule
\textbf{Setting} & \textbf{Model} & \textbf{Input} & \textbf{mAP} ($\uparrow$) & \textbf{ER$_{20^{\circ}}$} ($\downarrow$) & \textbf{MAE} ($\downarrow$) & \textbf{DER} ($\downarrow$) \\
\midrule
\multirow{4}{*}{\textbf{Single-source}}
 & AudioMAE~\cite{AudioMAE} & Mel-spectrograms (mono) & 47.18 & -    & -    & -    \\
 & SELDnet~\cite{SELDnet}   & Mel-spectrograms, IPD   & 42.66 & 25.19 & 19.21 & 38.46 \\
 & Spatial-AST~\cite{bat}   & Mel-spectrograms, IPD   & 49.86 & \textbf{23.97} & \textbf{18.03} & \textbf{32.96} \\
 & \textbf{DSAT (ours)}     & Mel-spectrograms, IPD   & \textbf{50.29} & 26.97 & 20.87 & 37.99 \\
\midrule
\multirow{2}{*}{\textbf{Dual-source: one source matched}}
 & Spatial-AST~\cite{bat}   & Mel-spectrograms, IPD   & 24.14 & 35.86 & 26.72 & 37.76 \\
 & \textbf{DSAT (ours)}     & Mel-spectrograms, IPD   & \textbf{31.35} & \textbf{27.29} & \textbf{22.27} & \textbf{30.41} \\
 \midrule
\multirow{2}{*}{\textbf{Dual-source: both sources matched}}
 & Spatial-AST~\cite{bat}   & Mel-spectrograms, IPD   & 15.52 & 64.75 & 54.45 & 46.64 \\
 & \textbf{DSAT (ours)}     & Mel-spectrograms, IPD   & \textbf{31.66} & \textbf{37.73} & \textbf{29.59} & \textbf{36.52} \\
\bottomrule
\label{Table:Encoder_seld}
\end{tabular*}
\end{table*}

\vspace*{-0.2cm}
\subsection{Baselines and Evaluation Metrics}
\textbf{Baselines.} For the spatial audio encoder, we compare DSAT against Spatial-AST~\cite{bat}, the encoder of BAT and the closest prior work, in both single-source and two-source settings, and against SELDnet~\cite{SELDnet} in the single-source setting. Following the protocol of~\cite{bat}, SELDnet's feature extraction network is augmented with a $12$-layer Transformer so that all compared encoders have approximately $90$M parameters. We additionally report AudioMAE~\cite{AudioMAE} as a monaural reference. Since it accepts only single-channel input, it reflects detection performance without access to binaural spatial cues and is therefore excluded from the spatial metrics. For the full model, we compare SAIL against BAT~\cite{bat} on SpatialSoundQA under the same three-stage curriculum. All BAT results are obtained by evaluating its officially released checkpoint\footnote{\url{https://github.com/X-LANCE/SLAM-LLM/tree/main/examples/seld_spatialsoundqa}} using our evaluation pipeline.

\textbf{Evaluation Metrics.}
We adopt the evaluation protocol of BAT~\cite{bat}. For the encoder, we report mean average precision (mAP) for sound event detection, DoA Error Rate at $20^{\circ}$ (ER$_{20^{\circ}}$) and Mean Angular Error (MAE) for DoA estimation, and Distance Error Rate (DER), defined as the fraction of predictions that deviate by more than $0.5$~m from the ground truth. For QA evaluation on SpatialSoundQA, we report mAP for detection questions, accuracy (Acc) for DoA questions, where the three-dimensional space is divided into eight directional regions, DER for distance questions, and Binary Accuracy (BA) for Yes/No reasoning questions. In addition to binary reasoning evaluation, we evaluate open-ended reasoning questions, reporting mAP for source-identification questions, Acc for relative-direction questions, and DER for relative-distance questions.

\vspace*{-0.1cm}
\subsection{Performance of the DSAT Encoder on SELD Tasks}
Table~\ref{Table:Encoder_seld} reports DSAT on single-source and two-source SELD. We compare against Spatial-AST~\cite{bat} in both settings, and against AudioMAE~\cite{AudioMAE} and SELDnet~\cite{SELDnet} in the single-source setting. Detection is measured by mAP, direction of arrival by ER$_{20^{\circ}}$ and MAE, and distance by DER. AudioMAE takes only monaural Mel-spectrograms and is therefore omitted from the spatial metrics.

\textbf{Single-source.}
On single-source clips, DSAT and Spatial-AST are comparable on detection ($50.29$ vs.\ $49.86$ mAP) and both clearly outperform AudioMAE ($47.18$) and SELDnet ($42.66$). Spatial-AST remains the strongest spatial estimator in this setting, with $23.97$ ER$_{20^{\circ}}$, $18.03$ MAE, and $32.96$ DER, while DSAT obtains $26.97$, $20.87$, and $37.99$, respectively. Relative to SELDnet, DSAT improves detection and is comparable on distance ($37.99$ vs.\ $38.46$ DER), but does not reduce angular error. This pattern is consistent with the design of DSAT: rather than specializing in a single aggregated prediction, it allocates capacity to source-discriminative query slots for multi-source spatial modelling.

\textbf{Two-source.} The advantage of DSAT becomes clear in two-source mixtures. Since Spatial-AST has only a single set of task queries and produces one prediction per clip regardless of the number of sources, we evaluate both models under two settings. In the ``one source matched" setting, each model is scored only on its best-covered source: the single prediction of Spatial-AST is matched to the closer ground-truth source by the spatial prediction costs, and correspondingly only the better-matched slot of DSAT is scored. In the ``both sources matched" setting, both ground-truth sources are counted: the two slots of DSAT are each scored against their assigned source, while the single prediction of Spatial-AST is scored against both sources, so that the two models are pooled over the same number of evaluation points. In the one-source-matched setting, DSAT substantially outperforms Spatial-AST, improving mAP from $24.14$ to $31.35$ and reducing ER$_{20^{\circ}}$ from $35.86$ to $27.29$, MAE from $26.72$ to $22.27$, and DER from $37.76$ to $30.41$. In the both-sources-matched setting, the gap widens further: DSAT improves mAP from $15.52$ to $31.66$ and reduces ER$_{20^{\circ}}$, MAE, and DER from $64.75$, $54.45$, and $46.64$ to $37.73$, $29.59$, and $36.52$, respectively.

Two observations support this. First, even when scored only on its best-covered source, Spatial-AST degrades markedly relative to its single-source performance (e.g., ER$_{20^{\circ}}$ rises from $23.97$ to $35.86$), indicating that the interfering source corrupts its single shared prediction. DSAT instead maintains stable per-source DoA accuracy (from $26.97$ to $27.29$ ER$_{20^{\circ}}$), as each source-discriminative query slot attends to its own target. Second, the gap between the two evaluation settings reveals how much of each model's performance covers both sources: DSAT loses little when both sources are counted (mAP $31.35\rightarrow31.66$), showing that both slots produce accurate predictions, whereas Spatial-AST drops sharply (mAP $24.14\rightarrow15.52$), since its single prediction can describe at most one source and the second is structurally missed. These results show that DSAT has a clear advantage over Spatial-AST on two-source SELD tasks. The source-discriminative query design and dual-source pre-training enable DSAT to produce more reliable source-level acoustic and spatial features, providing a stronger foundation for subsequent LLM reasoning over sound events, locations, and inter-source spatial relationships.

\begin{figure*}[!t]
\centering
\includegraphics[width=\linewidth]{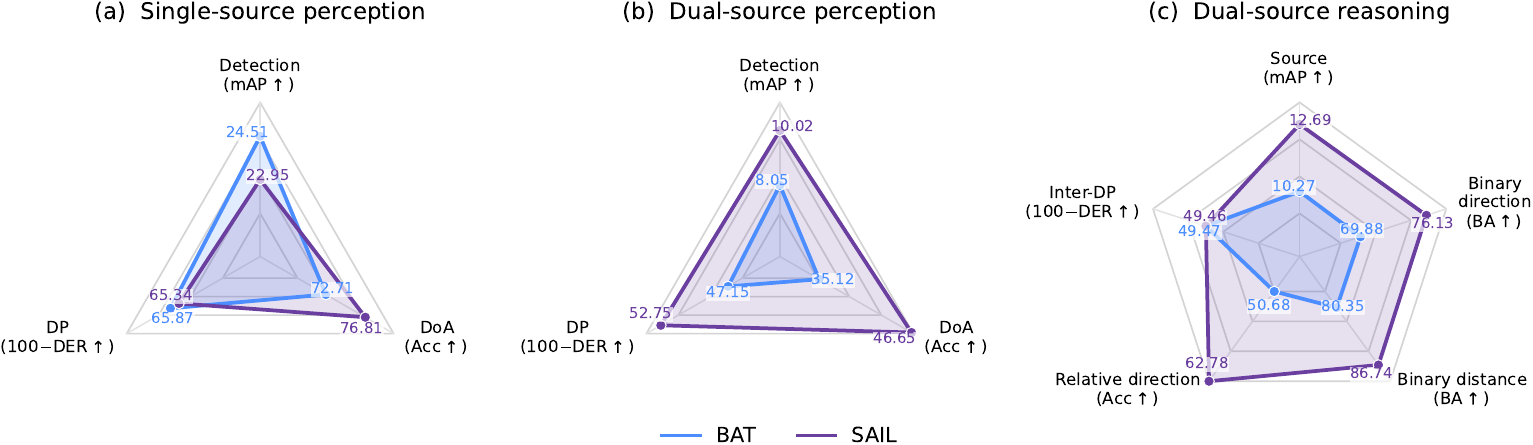}
\caption{Comparison of SAIL and BAT on the SpatialSoundQA benchmark for (a) single-source perception, (b) dual-source perception, and (c) dual-source reasoning. Each axis is independently normalized for visualization, while the annotations show the original metric values.}
\label{Fig SAIL BAT Radar}
\end{figure*}

\begin{table*}[t]
\centering
\begin{threeparttable}
\caption{Comparison of SAIL with BAT on SpatialSoundQA. Perception is evaluated on sound detection, DoA estimation, and distance prediction (DP). Values before and after ``$|$'' denote single-source (Types A and B) and dual-source (Types C and D) results, respectively, with single-source results shaded in grey. Type-E reasoning includes binary Yes/No questions evaluated by Binary Accuracy (Acc) and additional open-ended dual-source questions on direction-conditioned source identification (Source, mAP), relative source direction (Direction, Acc), and inter-source distance (DP, DER).}
\label{Table:SAIL}
\setlength{\tabcolsep}{4.5pt}
\begin{tabular}{llccccccccc}
\toprule
&& \multicolumn{3}{c}{Perception (Type ABCD)}
& \multicolumn{3}{c}{Reasoning (Type E, Binary Acc $\uparrow$)}
& \multicolumn{3}{c}{Open-ended Reasoning\tnote{$\ddagger$}} \\
\cmidrule(lr){3-5} \cmidrule(lr){6-8} \cmidrule(lr){9-11}
Model & Input
& \makecell{Detection \\ (mAP $\uparrow$)}
& \makecell{DoA \\ (Acc $\uparrow$)}
& \makecell{DP \\ (DER $\downarrow$)}
& Direction & Distance & Avg.
& \makecell{Source \\ (mAP $\uparrow$)}
& \makecell{Direction \\ (Acc $\uparrow$)}
& \makecell{Inter-source DP \\ (DER $\downarrow$)} \\
\midrule
Random & --
& \g{0.65} $|$ 0.64 & \g{12.69} $|$ 12.82 & \g{65.53} $|$ 77.78
& 49.99 & 49.66 & 49.83
& 0.77 & 50.28 & 63.79 \\
BAT~\cite{bat} & B + P
& \g{24.51} $|$ 8.05 & \g{72.71} $|$ 35.12 & \g{34.13} $|$ 52.85
& 69.88 & 80.35 & 75.12
& 10.27 & 50.68 & \textbf{50.53} \\
\midrule
\multirow{2}{*}{SAIL (ours)} & P
& \g{0.61} $|$ 0.60 & \g{14.34} $|$ 13.62 & \g{59.26} $|$ 71.63
& 48.06 & 53.41 & 50.74
& 0.79 & 46.38 & 89.14 \\
 & B + P
& \g{\textbf{22.95}} $|$ \textbf{10.02} & \g{\textbf{76.81}} $|$ \textbf{46.65} & \g{\textbf{34.66}} $|$ \textbf{47.25}
& \textbf{76.13} & \textbf{86.74} & \textbf{81.44}
& \textbf{12.69} & \textbf{62.78} & 50.54 \\
\bottomrule
\end{tabular}
\begin{tablenotes}\footnotesize
\item Input types: B = binaural audio, P = text prompt. The "SAIL (P)" row receives no audio and serves as a language-prior control.
\item Random baselines for this subset are computed by us via sampling from the answer distribution.
\item[$\ddagger$] Open-ended questions all involve dual-source audio and require the model to directly produce the answer rather than verify a given proposition.
\end{tablenotes}
\end{threeparttable}
\end{table*}

\subsection{Spatial Audio Reasoning}
Table~\ref{Table:SAIL} compares SAIL with BAT~\cite{bat} on the SpatialSoundQA benchmark, while Fig.~\ref{Fig SAIL BAT Radar} visualizes their performance across single-source perception, dual-source perception, and dual-source reasoning tasks. Overall, SAIL achieves performance comparable to BAT on single-source perception and demonstrates clear advantages in dual-source perception and spatial reasoning.

\textbf{Spatial audio perception.}
For single-source perception, SAIL and BAT achieve comparable overall performance. SAIL improves DoA accuracy from $72.71$ to $76.81$, while BAT performs slightly better in sound detection ($24.51$ vs.\ $22.95$ mAP) and distance prediction ($34.13$ vs.\ $34.66$ DER). The advantage of SAIL becomes more evident in dual-source scenarios. Compared with BAT, SAIL improves detection mAP from $8.05$ to $10.02$ and DoA accuracy from $35.12$ to $46.65$, while reducing DER from $52.85$ to $47.25$. In particular, the $32.83\%$ relative improvement in dual-source DoA accuracy indicates that SAIL can more effectively distinguish and localize individual sources in overlapping acoustic scenes. These results are consistent with the source-discriminative design of DSAT, which produces source-level representations instead of encoding a two-source mixture into a single shared prediction.

\textbf{Binary spatial reasoning.}
SAIL consistently outperforms BAT on the binary Type-E reasoning tasks. For direction reasoning, SAIL improves accuracy from $69.88$ to $76.13$, while distance reasoning accuracy increases from $80.35$ to $86.74$. Consequently, the average binary reasoning accuracy increases from $75.12$ to $81.44$, representing a relative improvement of $8.41\%$. As discussed in BAT~\cite{bat}, these tasks require the model to identify the two overlapping sound sources, estimate their spatial properties, and reason about their relationships in the context of a natural-language question. The consistent improvements on both direction and distance reasoning suggest that the source-level acoustic and spatial representations produced by DSAT provide the LLM with more reliable information for multi-source reasoning.

\textbf{Open-ended spatial reasoning.}
We further evaluate the models using open-ended questions that require them to directly generate source identities and spatial relationships rather than verify a given proposition. SAIL improves direction-conditioned source identification from $10.27$ to $12.69$ in mAP and relative-direction accuracy from $50.68$ to $62.78$. This substantial improvement in relative-direction reasoning shows that SAIL more effectively associates each sound event with its corresponding spatial location. For inter-source distance prediction, SAIL and BAT perform almost identically, with DERs of $50.54$ and $50.53$, respectively, indicating that accurately estimating the distance between two concurrent sources remains challenging.

The prompt-only SAIL control performs close to the random baseline, achieving an average binary reasoning accuracy of $50.74$ compared with $49.83$ for random prediction. It also remains near chance on most perception and open-ended reasoning metrics. This confirms that SAIL's improvements do not arise primarily from language priors in the questions and instead depend on information extracted from the binaural audio. Therefore, the results demonstrate that SAIL preserves competitive single-source perception while substantially improving dual-source perception and spatial reasoning, highlighting the importance of source-discriminative spatial audio representations for reasoning about complex acoustic scenes.

\begin{table*}[t]
\centering
\caption{Ablation on the LLM backbone of SAIL. All variants share the same DSAT encoder, Q-Former, and three-stage training curriculum, differing only in the underlying LLM. \#Params denotes the total parameter count of the LLM backbone. Metrics and column conventions follow Table~\ref{Table:SAIL}: values before and after ``$|$'' correspond to single-source and dual-source questions, respectively, with single-source entries shaded in grey.}
\label{tab:backbone}
\setlength{\tabcolsep}{4pt}
\begin{tabular}{llccccccccc}
\toprule
&& \multicolumn{3}{c}{Perception (Type ABCD)}
& \multicolumn{3}{c}{Reasoning (Type E, Binary Acc $\uparrow$)}
& \multicolumn{3}{c}{Open-ended Reasoning} \\
\cmidrule(lr){3-5} \cmidrule(lr){6-8} \cmidrule(lr){9-11}
LLM Backbone & \#Params
& \makecell{Detection \\ (mAP $\uparrow$)}
& \makecell{DoA \\ (Acc $\uparrow$)}
& \makecell{DP \\ (DER $\downarrow$)}
& Direction & Distance & Avg.
& \makecell{Source \\ (mAP $\uparrow$)}
& \makecell{Direction \\ (Acc $\uparrow$)}
& \makecell{Inter-source DP \\ (DER $\downarrow$)} \\
\midrule
Llama-2-7b & 7B
& \g{22.95} $|$ 10.02 & \g{76.81} $|$ 46.65 & \g{34.66} $|$ 47.25 & 76.13 & 86.74 & 81.44 & 12.69 & 62.78 & 50.54 \\
Llama-3.1-8B & 8B
& \g{23.11} $|$ 9.01 & \g{76.71} $|$ 45.67 & \g{33.54} $|$ 47.53 & 74.24 & 82.48 & 78.36 & 11.78 & 60.30 & 47.71\\
\bottomrule
\end{tabular}
\end{table*}

\subsection{Compatibility with Different LLM Backbones}
\label{sec:ablation-backbone}
Table~\ref{tab:backbone} examines the effect of using different LLM backbones while keeping the DSAT encoder, Dual-Stream Q-Former, and training curriculum unchanged. The two variants show comparable perception performance, particularly for direction and distance estimation. This indicates that the spatial information is mainly captured by the DSAT encoder and transferred through the Q-Former, making the perception results relatively insensitive to the choice of LLM.

Llama-2-7B \cite{Llama2} performs better on most reasoning metrics despite having fewer parameters. It achieves an average binary reasoning accuracy of 81.44\%, compared with 78.36\% for Llama-3.1-8B \cite{llama3}, with improvements in both direction and distance reasoning. It also performs better in open-ended source identification and direction reasoning, while Llama-3.1-8B obtains a lower inter-source distance error. These results indicate that a newer or larger language backbone does not necessarily improve spatial audio reasoning, since performance also depends on the alignment between the audio representations and the LLM. Considering its stronger overall reasoning performance and smaller parameter count, we adopt Llama-2-7B as the default backbone of SAIL.

\begin{figure}[!t]
\centering
\includegraphics[width=\linewidth]{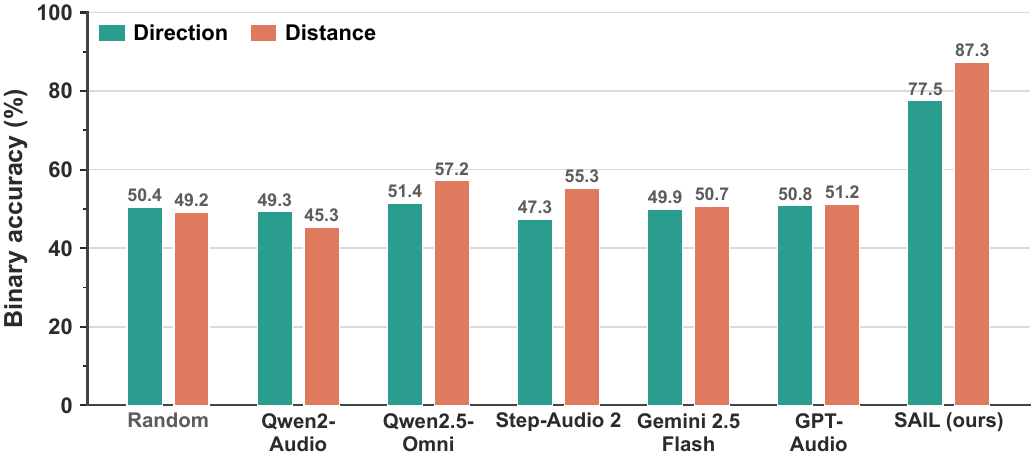}
\caption{Comparison of SAIL and general-purpose audio LLMs using dual-source binary direction and distance reasoning questions. All models are evaluated on the same set of $1{,}000$ instances randomly sampled from SpatialSoundQA.}
\label{Fig llm-binary comparison}
\end{figure}

\begin{figure*}[!t]
\centering
\includegraphics[width=\linewidth]{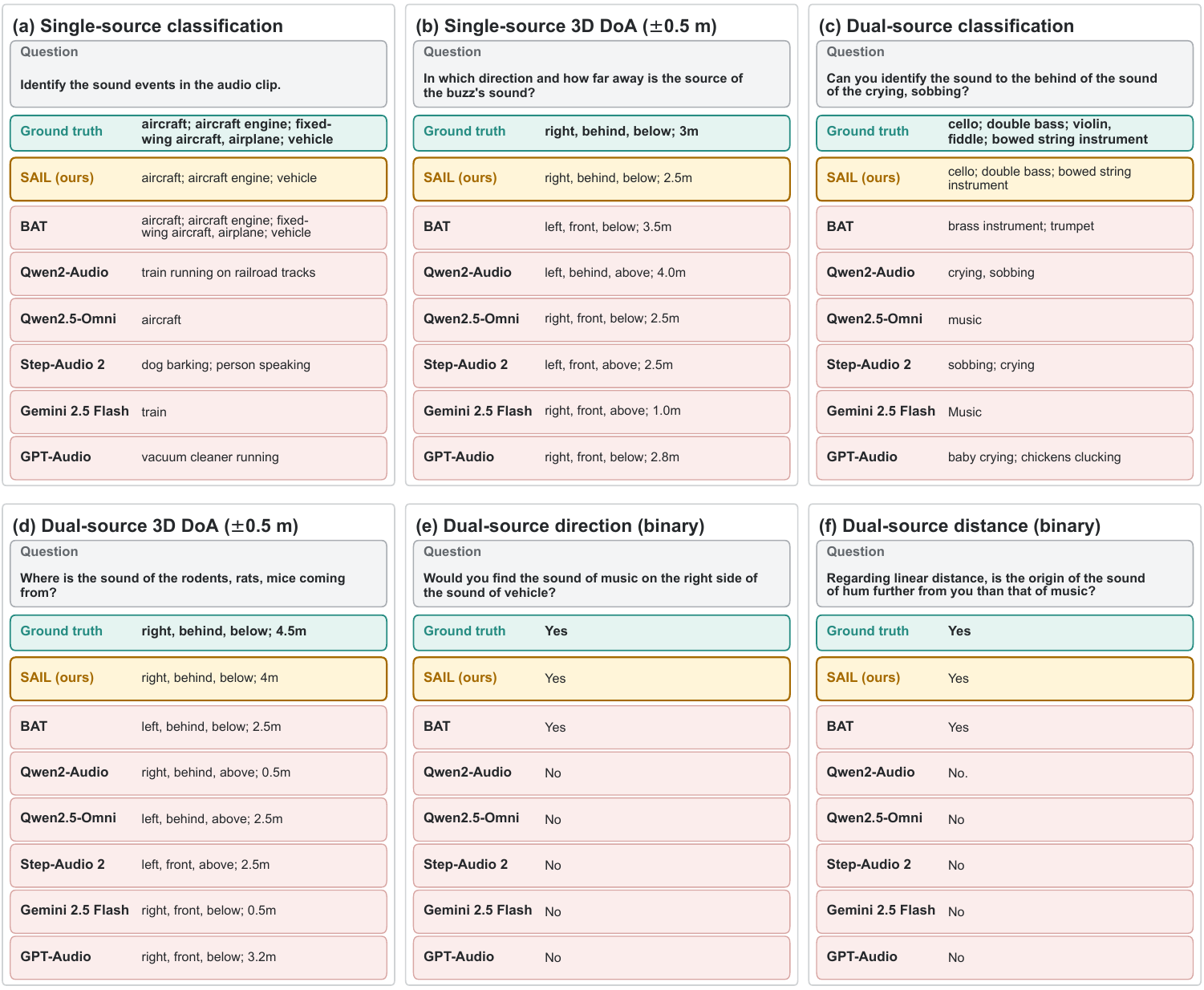}\vspace*{-0.1cm}
\caption{Qualitative comparison of SAIL, BAT, and general-purpose audio LLMs on SpatialSoundQA. The examples cover (a) single-source classification, (b) single-source 3D direction-of-arrival (DoA) estimation, (c) dual-source classification, (d) dual-source 3D DoA estimation, (e) dual-source binary direction reasoning, and (f) dual-source binary distance reasoning. For DoA estimation, distance predictions within $\pm0.5$\,m of the ground truth are considered correct.}
\label{Fig SAIL_qualitative_examples}
\end{figure*}

\subsection{Comparison with General-Purpose Audio LLMs}
Fig.~\ref{Fig llm-binary comparison} compares SAIL against general-purpose audio LLMs on 1,000 dual-source binary spatial reasoning questions sampled from SpatialSoundQA~\cite{bat}. All models receive the same binaural audio files and text prompts. For proprietary models, the original two-channel files are submitted through their public APIs; whether the two channels are preserved is determined by each model's internal preprocessing pipeline. SAIL achieves 77.5\% accuracy for direction reasoning and 87.3\% for distance reasoning, substantially outperforming all evaluated general-purpose models. Their direction accuracy ranges from 47.3\% to 51.4\%, close to the random-guessing baseline of 50.4\%. Although their distance accuracy varies more widely from 45.3\% to 57.2\%, it remains considerably below SAIL.

The near-chance performance of general-purpose audio LLMs primarily reflects a mismatch between their audio front ends and binaural spatial reasoning, rather than a limitation of language reasoning alone. Binaural localization relies on interaural time, phase, and level differences between the two channels; however, these cues are often lost when multi-channel recordings are converted into a monaural waveform. For example, the Gemini API~\cite{Gemini2.5} explicitly combines multiple input channels into a single channel, while STAR-Bench~\cite{starbench} empirically identifies channel averaging as a common bottleneck that prevents existing audio LLMs from exploiting genuine stereo cues. This limitation affects direction reasoning most directly. Distance reasoning can retain partial monaural cues, such as propagation attenuation and the direct-to-reverberant energy ratio, which may explain the moderately above-chance results of Qwen2.5-Omni \cite{Qwen2.5-Omni} and Step-Audio~2 \cite{Step-audio2}. Nevertheless, comparing two concurrent sources also requires source separation, source-attribute binding, and relational reasoning. These capabilities are not explicitly targeted by general-purpose models trained primarily for general audio question answering. In contrast, SAIL preserves binaural cues and explicitly learns source-specific associations between sound events and their spatial attributes.

The qualitative examples in Fig.~\ref{Fig SAIL_qualitative_examples} further compare the models under single-source and dual-source conditions. In the single-source examples, SAIL recognizes most sound events and correctly estimates all directional components, with a distance error of only 0.5\,m. BAT performs well in sound classification but incorrectly predicts the source direction, while the general-purpose models show partial recognition and errors in multiple spatial dimensions. In the more challenging dual-source examples, SAIL successfully distinguishes the concurrent events and associates the queried source with its location, whereas BAT and the general-purpose models exhibit more evident source confusion and localization errors. For the binary questions, both SAIL and BAT correctly infer the relative direction and distance, while all general-purpose models give the opposite answers, demonstrating the benefit of spatial audio training.

\vspace*{-0.3cm}
\section{Limitations and Future Work}
First, although SAIL is formulated for a general number of source slots, the current model is trained and evaluated only on scenes containing at most two sources. Extending it to more sources requires additional source slots, causing the number of audio tokens and the associated LLM computation to grow with the number of sources. Adaptive slot allocation or token compression may therefore be necessary for efficient reasoning in many-source scenes.

Second, the current front end is specifically designed for binaural audio. The IPD features are derived from a single channel pair, while the Mel branch is configured for two-channel inputs. Extending SAIL to richer spatial audio formats, such as Ambisonics or microphone arrays, would require a spatial encoding mechanism capable of modeling more general inter-channel relationships. We plan to explore such multi-channel spatial ALMs in future work.

\section{Conclusion}\label{Conclusion}
This work presented SAIL, a spatial ALM designed to preserve acoustic, spatial, and source-level structure throughout audio encoding and language alignment. SAIL introduces the DSAT encoder to model Mel and IPD features as disentangled token streams, source-discriminative task queries to learn source-level representations, and a Dual-Stream Q-Former to produce structured acoustic and spatial tokens for the LLM. Experiments on SpatialSoundQA show that SAIL maintains competitive single-source perception while substantially improving dual-source detection, localization, and spatial reasoning over BAT. It also clearly outperforms general-purpose ALMs for direction and distance reasoning. Compared with the early-fusion baseline, SAIL improves mAP for two-source detection from 8.05 to 10.02, increases direction accuracy from 35.12\% to 46.65\%, and reduces the distance error rate from 52.85\% to 47.25\%. It also improves average binary spatial reasoning accuracy from 75.12\% to 81.44\%. These results demonstrate the importance of preserving binaural cues and source-location correspondence, and structured, source-discriminative audio representations for multi-source spatial perception and reasoning. Future work will extend SAIL to scenes with more sound sources, stronger source separation, and more general multichannel audio formats.

\section*{Acknowledgment}
We thank the authors of ``BAT: Learning to Reason about Spatial Sounds with Large Language Models''~\cite{bat} for making their code and pretrained model checkpoints publicly available.

\appendices
\section{Training Configurations}\label{Training Configurations}
Table~\ref{Table Training configurations} summarizes the training configurations of the DSAT encoder and SAIL. DSAT is trained progressively from single-source detection to mixed-source spatial perception. SAIL subsequently follows a three-stage curriculum while keeping the DSAT encoder and base LLM frozen and updating only the Dual-Stream Q-Former projector and LoRA adapters. The learning rates, warmup settings, batch sizes, and loss weights used at each stage are also reported for reproducibility.

\begin{table*}[t]
\centering
\caption{Training configurations for the DSAT encoder and SAIL.}
\setlength{\tabcolsep}{3.5pt}
\resizebox{\textwidth}{!}{
\begin{tabular}{lcccccc}
\toprule
& \multicolumn{3}{c}{DSAT Encoder Pre-training}
& \multicolumn{3}{c}{SAIL Instruction Tuning} \\
\cmidrule(lr){2-4}\cmidrule(lr){5-7}
Configuration
& \makecell{Single-source\\Stage I}
& \makecell{Single-source\\Stage II}
& \makecell{Mixed-source}
& Stage I & Stage II & Stage III \\
\midrule
Training objective
& Detection
& \makecell{Detection,\\distance, DoA}
& \makecell{Detection,\\distance, DoA}
& \makecell{Single-source\\perception}
& \makecell{Dual-source\\perception}
& \makecell{Spatial\\reasoning} \\
\midrule
Trainable modules
& DSAT
& DSAT
& DSAT
& \multicolumn{3}{c}{Dual-Stream Q-Former and LoRA} \\
\midrule
Training loss
& Detection BCE
& Multi-task
& Multi-task
& \multicolumn{3}{c}{Language-modeling cross-entropy} \\
\midrule
$\lambda_{\mathrm{cls}}:\lambda_{\mathrm{dist}}:\lambda_{\mathrm{doa}}$
& $1000:0:0$
& $400:4:2$
& $400:8:2$
& \multicolumn{3}{c}{Not applicable} \\
\midrule
Initialization
& AudioMAE
& Previous stage
& Single-source DSAT
& Random initialization
& SAIL Stage I
& SAIL Stage II \\
\midrule
Optimizer
& \multicolumn{6}{c}{AdamW} \\
\midrule
Base Learning rate
& $1\times10^{-3}$
& $2\times10^{-4}$
& $2\times10^{-4}$
& $2\times10^{-4}$
& $2\times10^{-4}$
& $2\times10^{-4}$ \\
\midrule
Learning rate schedule
& \multicolumn{3}{c}{Linear warmup and cosine decay}
& \multicolumn{3}{c}{Linear warmup and linear decay} \\
\midrule
Warmup
& 3 epochs & 3 epochs & 3 epochs
& $\sim4.4\%$ steps & $\sim4.4\%$ steps & $\sim4.4\%$ steps \\
\midrule
Epochs
& 20 & 10 & 20 & 3 & 3 & 3 \\
\midrule
Per-GPU batch size
& 128 & 128 & 128
& 32 & 24 & 24 \\
\bottomrule
\end{tabular}}
\label{Table Training configurations}
\end{table*}

\bibliographystyle{IEEEtran}
\bibliography{main}

\end{document}